\documentclass[runningheads]{llncs}
\usepackage{graphicx}
\usepackage{comment}
\usepackage{amsmath,amssymb} % define this before the line numbering.
\usepackage{color}
\usepackage{caption}
\usepackage{epsfig}
\usepackage{tabularx}
\usepackage{microtype}
\usepackage{array}
\usepackage{booktabs}
\usepackage{dcolumn}
\usepackage{lipsum}
\usepackage{multicol}
\usepackage{multirow}
\usepackage{booktabs}
\usepackage{arydshln}
\usepackage{tabulary}
\usepackage{floatrow}
\usepackage{wrapfig}
\usepackage{sidecap}

\begin{document}
% \linenumbers
\pagestyle{headings}
\mainmatter
\def\ECCVSubNumber{1421}  % Insert your submission number here

\title{PIPAL: a Large-Scale Image Quality Assessment Dataset for Perceptual Image Restoration} % Replace with your title

\titlerunning{PIPAL Dataset for Perceptual Image Restoration}
\authorrunning{Gu et al.}
%%%%%%%%%%%
\author{Jinjin Gu\inst{1},
Haoming Cai\inst{1,2},
Haoyu Chen\inst{1},
Xiaoxing Ye\inst{1},\\
Jimmy S. Ren\inst{3}, and
Chao Dong\inst{2,4}}
%%%%%%%%%%%%%%
\institute{
The School of Data Science, The Chinese University of Hong Kong, Shenzhen\and
ShenZhen Key Lab of Computer Vision and Pattern Recognition, SIAT-SenseTime Joint Lab, Shenzhen Institutes of Advanced Technology, Chinese Academy of Sciences\and
SenseTime Research\and
SIAT Branch, Shenzhen Institute of Artificial Intelligence and Robotics for Society\\
\email{\{jinjingu, haomingcai, haoyuchen, xiaoxingye\}@link.cuhk.edu.cn}\\
\email{rensijie@sensetime.com, chao.dong@siat.ac.cn}}
%******************
\maketitle

\begin{abstract}
Image quality assessment (IQA) is the key factor for the fast development of image restoration (IR) algorithms.
The most recent IR methods based on Generative Adversarial Networks (GANs) have achieved significant improvement in visual performance, but also presented great challenges for quantitative evaluation.
Notably, we observe an increasing inconsistency between perceptual quality and the evaluation results.
Then we raise two questions:
(1) Can existing IQA methods objectively evaluate recent IR algorithms?
(2) When focus on beating current benchmarks, are we getting better IR algorithms?
To answer these questions and promote the development of IQA methods, we contribute a large-scale IQA dataset, called Perceptual Image Processing Algorithms (PIPAL) dataset.
Especially, this dataset includes the results of GAN-based methods, which are missing in previous datasets.
We collect more than 1.13 million human judgements to assign subjective scores for PIPAL images using the more reliable ``Elo system''.
Based on PIPAL, we present new benchmarks for both IQA and super-resolution methods.
Our results indicate that existing IQA methods cannot fairly evaluate GAN-based IR algorithms.
While using appropriate evaluation methods is important, IQA methods should also be updated along with the development of IR algorithms.
At last, we improve the performance of IQA networks on GAN-based distortions by introducing anti-aliasing pooling.
Experiments show the effectiveness of the proposed method.

\keywords{Perceptual Image Restoration, Image Quality Assessment, Generative Adversarial Network, Perceptual Super-Resolution}

\end{abstract}

\section{Introduction}
\label{sec:introduction}
Image restoration (IR) is a classic low-level vision problem that aims to reconstruct high-quality images from distorted low-quality inputs.
Typical IR tasks include image super-resolution (SR), denoising, enhancement, etc.
The whirlwind of deep-learning progress has produced a steady stream of promising IR algorithms that could generate less-distorted or perceptual-friendly images.
Nevertheless, one of the key bottlenecks that restrict IR methods' future development is the ``evaluation mechanism''.
Although it is nearly effortless for human eyes to distinguish perceptually better images, it is challenging for an algorithm to measure visual quality fairly.
In this work, we will focus on the analysis of existing evaluation methods and introduce a new image quality assessment (IQA) dataset, which not only includes the most recent IR methods but also has the largest scale/diversity.
The motivation will be first stated as follows.

IR methods are generally evaluated by measuring the similarity between the reconstructed images and ground-truth images via IQA metrics, such as PSNR \cite{psnr} and SSIM \cite{ssim}.
Recently, some non-reference IQA methods, such as Ma \cite{ma2017learning} and Perceptual Index (PI) \cite{blau2018perception}, are introduced to evaluate the recent perceptual-oriented algorithms. 
To some extent, these IQA methods are the chief reason for the considerable progress of the IR field.
However, while new algorithms have been continuously improving IR performance, we notice an increasing inconsistency between quantitative results and perceptual quality.
For example, literature \cite{blau2018perception} reveals that the superiority of PSNR values does not always accord with better visual quality. 
Although Blua \textit{et al.} suggest that PI is more relevant to human judgement, algorithms with high PI scores (e.g., ESRGAN \cite{wang2018esrgan} and RankSRGAN \cite{zhang2019ranksrgan}) could still produce images with obvious unrealistic artifacts.
These conflicts lead us to rethink the evaluation methods for IR tasks.

An important reason for this situation is the invention of Generative Adversarial Networks (GANs) \cite{goodfellow2014generative} and GAN-based IR methods \cite{wang2018esrgan,gu2019image}, bringing completely new characteristics to the output images.
In general, these methods often fabricate seemingly realistic yet fake details and textures.
This presents a great challenge for existing IQA methods, which cannot distinguish the GAN-generated textures from noises and real details.
We naturally raise two questions:
(1) Can existing IQA methods objectively evaluate current IR methods, especially GAN-based methods?
(2) With the focus on beating benchmarks on the flawed IQA methods, are we getting better IR algorithms?
A few works have made early attempt to answer these questions by proposing new benchmarks for IR and IQA methods.
Yang \textit{et al.} \cite{yang2014single} conduct a comprehensive evaluation of traditional SR algorithms.
Blau \textit{et al.} \cite{blau2018perception} analyze the perception-distortion trade-off phenomenon and suggest the use of multiple IQA methods.
However, these prior studies usually apply unreliable human ratings of image quality, and are generally insufficient in the number of IR/IQA methods.
Especially, the results of GAN-based methods are missing in the above works.

To touch the heart of this problem, we need to have a better understanding of the new challenges brought by GAN.
The first issue is to build a new IQA dataset with GAN-based algorithms.
In this work, we contribute a novel IQA dataset, namely Perceptual Image Processing ALgorithms dataset (PIPAL).
This IQA dataset includes a lot of distorted images with visual quality levels annotated by humans.
It can be used to measure the consistency of the prediction of IQA method and human judgement.
The proposed PIPAL dataset distinguishes from previous datasets in three aspects:
(1) In addition to traditional distortion types (e.g., Gaussian noise/blur), PIPAL contains the outputs of several kinds of IR algorithms, including traditional algorithms, deep-learning-based algorithms and GAN-based algorithms. In particular, this is the first time for the results of GAN-based algorithms to appear in an IQA dataset.
(2) We employ the Elo rating system \cite{elo1978rating} to assign subjective scores, involving more than 1.13 million human judgements. Comparing with existing rating systems (e.g., five gradations \cite{live} and Swiss system \cite{tid2008}), the Elo rating system provides much more reliable probability-based rating results. Furthermore, it has good extensibility, allowing users to update the dataset by directly adding new distortion types.
(3) The proposed dataset contains 29k images in total, including 250 high-quality reference images, and each of which has 116 distortions. To date, PIPAL is the largest IQA dataset with complete subjective scoring.

With the PIPAL dataset, we are able to answer the questions above.
(1) We build a benchmark using the proposed PIPAL dataset for existing IQA methods.
Experiments indicate that PIPAL poses challenges for these IQA methods.
Evaluating IR algorithms only using existing metrics is not appropriate.
Our research also shows that compared with the widely-used metrics (e.g., PSNR and PI), PieAPP \cite{prashnani2018pieapp} and LPIPS \cite{zhang2018unreasonable} are more suitable for evaluating IR algorithms, especially GAN-based algorithms.
(2) We then review the development of SR algorithms in recent years.
The results show that the recent SR algorithms achieve great progress in the average subjective image quality scores.
However, we find that none of the existing IQA methods is always effective in evaluating SR algorithms.
With the invention of new IR technologies, the corresponding evaluation methods also need to be adjusted to continuously promote the development of the IR field.
(3) We also study the characteristics of GAN-based distortion by comparing them with some well-studied traditional distortions.
Based on the results, we argue that existing IQA methods' low tolerance toward spatial misalignment may be one of the key reasons for their performance drop.
By introducing anti-aliasing pooling to the existing IQA networks, we are able to improve their performance on GAN-based distortions.

\section{Related Work}
\label{sec:related}
\paragraph{Image Restoration.}
As a fundamental computer vision problem, IR aims at recovering a high-quality image from its degraded observations.
In past decades, plenty of IR algorithms have been proposed to continuously improve the performance.
The early algorithms use hand-craft features \cite{bm3d,ywhm2010} or exploit image priors \cite{tsg2013,a+2014} in optimization problems to reconstruct images.
Since the pioneer work of using Convolution Neural Networks (CNNs) to learn the IR mappings \cite{jain2009natural,srcnn2014}, the deep-learning-based algorithms have dominated IR research due to their remarkable performance and usability \cite{feng2019suppressing,gu2019blind}.
Recently, with the invention of GAN \cite{goodfellow2014generative}, GAN-based IR methods \cite{enhancenet2017,zhang2019ranksrgan} are not limited to getting a higher PSNR performance but trying to have better perceptual effect.
However, these IR algorithms are not perfect.
The results of those algorithms also include various image defects, and they are different from the traditional distortions that are often discussed in previous IQA researches.
With the development of IR algorithms and the emergence of new technologies, evaluating the results of these algorithms becomes more and more challenging.
In this paper, we mainly focus on the restoration of low-resolution images, noisy images, and images degraded by both resolution reduction and noise.

\begin{table}[t!]
    \centering
    \resizebox{1.0\linewidth}{!}{
    \begin{tabular}{cccccccc}
        \toprule
        \multirow{2}{*}{\textbf{Dataset}}  &
        \# Ref. &
        Image &
        Distortion &
        \# Distort. &
        \# Distort. &
        \# Human &
        judgement \\
        & 
        images &
        types &
        types &
        types &
        images &
        judgements &
        type \\
        \toprule
        LIVE \cite{live} &
        29 &
        image &
        traditional &
        5 &
        0.8k &
        25k &
        MOS (Five gradations) \\
        CSIQ \cite{csiq} &
        30 &
        image &
        traditional &
        6 &
        0.8k &
        5k &
        MOS (Direct ranking) \\
        TID2008 \cite{tid2008} &
        25 &
        image &
        traditional &
        17 &
        1.7k &
        256k &
        MOS (Swiss system)\\
        TID2013 \cite{tid2013} &
        25 &
        image &
        traditional &
        24 &
        3.0k &
        524k &
        MOS (Swiss system) \\
        BAPPS* \cite{zhang2018unreasonable} &
        187.7k &
        patch (256$\times$256) &
        trad. $+$ alg. outputs&
        425 &
        375.4k &
        484.3k &
        Prob. of Preference \\
        PieAPP* \cite{prashnani2018pieapp} &
        200 &
        patch (256$\times$256) &
        trad. $+$ alg. outputs&
        75 &
        20.3k &
        2.3m &
        Prob. of Preference \\
        \hline
        PIPAL &
        \multirow{2}{*}{250} &
        patch &
        trad. $+$ alg. outputs&
        \multirow{2}{*}{40} &
        \multirow{2}{*}{29k} &
        \multirow{2}{*}{1.13m} &
        MOS \\
        (Ours)&
        &
        (288$\times$288)&
        \emph{including GAN} &
        &
        &
        &
        (Elo rating system)\\
        \toprule
    \end{tabular}
    }
    \captionsetup{font=small}
    \caption{Comparison with the previous datasets.
    We include the outputs of GAN-based algorithms as a novel distortion type. 
    Note that BAPPS \cite{zhang2018unreasonable} and PieAPP are perceptual similarity dataset (as opposed to IQA datasets), and are marked with ``*''}
    \label{tab:datasets}
\end{table}{}  % Dataset comparsion Table

\paragraph{Image Quality Assessment.}
The IQA methods were developed to measure the perceptual quality of images after degradation or post-processing operations.
According to different usage scenarios, IQA methods can be divided into full-reference methods (FR-IQA) and no-reference methods (NR-IQA).
FR-IQA methods measure the similarity between two images from the perspective of information or perceptual feature similarity, and have been widely used in the evaluation of image/video coding, restoration and communication quality.
Beyond the most widely-used PSNR, FR-IQA methods follow a long line of works that can trace back to SSIM \cite{ssim}, which first introduces structural information in measuring image similarity.
After that, various FR-IQA methods have been proposed to bridge the gap between the results of IQA methods and human judgements.
Similar to other computer vision problems, advanced data-driven methods have also motivated the investigation of applications of IQA \cite{zhang2018unreasonable,prashnani2018pieapp}.
In addition to the above FR-IQA methods, NR-IQA methods are proposed to assess image quality without a reference image.
Some popular NR-IQA methods include NIQE \cite{niqe}, Ma \textit{et al.} \cite{ma2017learning}, BRISQUE \cite{brisque}, and PI \cite{blau2018perception}.
In some recent works, NR-IQA and FR-IQA methods are combined to measure IR algorithms \cite{blau2018perception}.
Despite of the progress of IQA methods, only a few IQA methods (e.g., PSNR, SSIM and PI) are frequently used to evaluate IR methods.

\paragraph{Image Quality Assessment Datasets.}
In order to evaluate and develop IQA methods, many datasets have been proposed, such as LIVE \cite{live}, CSIQ \cite{csiq}, TID2008 and TID2013 \cite{tid2008,tid2013}.
There are also some perceptual similarity datasets such as PieAPP \cite{prashnani2018pieapp}, and BAPPS \cite{zhang2018unreasonable}.
These datasets provide both distorted images and the corresponding subjective scores, and they have served as baselines for evaluation of IQA methods.
These datasets are mainly distinguished from each other in three aspects: (1) the collecting of the reference images, (2) the number of distortions included and their types and (3) the collecting strategy of subjective score.
A quick comparison of these datasets can be found in \tablename~\ref{tab:datasets}.

\section{Perceptual Image Processing ALgorithms Dataset}
\label{sec:pipal}

We first describe the peculiarities of the proposed dataset from the aforementioned aspects of (1) the collecting of the reference images, (2) the number of distortions and their types, and (3) the collecting of subjective score, respectively.

\paragraph{Collection of reference images.}
In the proposed dataset, we select 250 image patches from two high-quality image datasets -- DIV2K \cite{div2k} and Flickr2K \cite{timofte2017ntire}.
We mainly focus on the area that is relatively hard to restore, such as high-frequency textures.
Thus, we crop patches of the representative texture areas from the selected images.
The selected reference images are representative of a wide variety of real-world textures, including but not limited to buildings, trees, grasses, animal fur, human faces, text, and artificial textures.
The size of these images is $288\times288$, which could meet the requirements of most IQA methods.

\begin{table}[t!]
    \centering
    \footnotesize
    \resizebox{1.0\linewidth}{!}{
    \begin{tabular}{p{3.6cm}<{\centering}p{11cm}<{\centering}}
        \toprule
        \textbf{Sub-type} & \textbf{Distortion Types} \\
        \hline
        Traditional & Gaussian blur, motion blur, image compression, Gaussian noise, spatial warping, bilateral filter, comfort noise.\\
        \hline
        Super-Resolution & interpolation method, traditional methods, SR with kernel mismatch, PSNR-oriented methods, GAN-based methods. \\
        \hline
        Denoising & mean filtering, traditional methods, deep-learning-based methods. \\
        \hline
        Mixture Restoration & SR of noisy images, SR after denoising, SR after compression noise removal \\
        \toprule
    \end{tabular}
    }
    \captionsetup{font=small}
    \caption{Our distortions types. In addition to the existing distortions, we include 19 different GAN-based algorithms distortions}
    \label{tab:distortions}
\end{table}{}  % Distortion types table

\begin{figure}[t]
    \centering
    \includegraphics[width=1.0\linewidth]{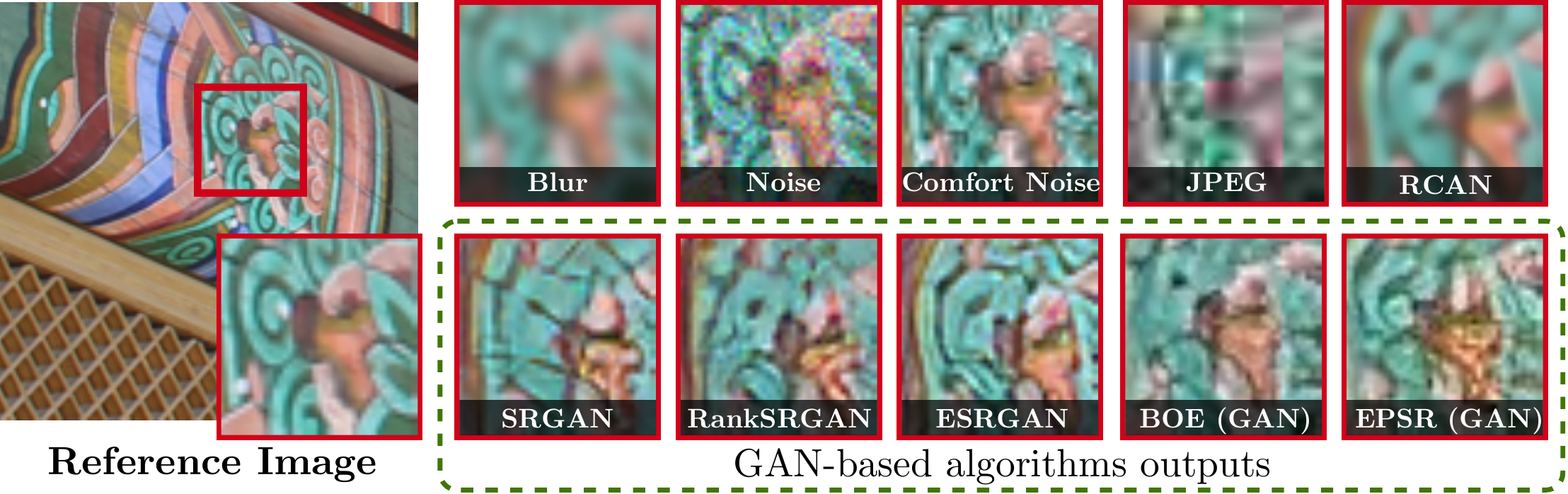}
    \captionsetup{font=small}
    \caption{Visualizing different distortions. Unlike the distortions in the upper row, which do not follow the natural image distribution. The GAN-based outputs are actually similar to natural images. However, their details are wrong}
    \label{fig:GANvis}
\end{figure}

\paragraph{Image Distortions.}
In our dataset, we have 40 distortion types, which can be divided into four sub-types.
An over-view of these distortion types is shown in \tablename~\ref{tab:distortions}.
The first sub-type includes some traditional distortions (e.g., blur, noise, and compression), which are usually performed by basic low-level image editing operations.
In some datasets, these distortions can be very severe; however, in our dataset, we constrain the situation of severe distortions as we want these distortions to be comparable to the IR results, which are not likely to be very low-quality.
The second sub-type includes the SR results from existing algorithms.
Although some recent datasets \cite{zhang2018unreasonable,prashnani2018pieapp} have covered some of the SR results, they contain results that are inferior in algorithms number and types to our dataset.
We divide the selected SR algorithms into three categories -- traditional algorithms, PSNR-oriented algorithms, and GAN-based algorithms.
The results of traditional algorithms can be understood, to some extent as, loss of detail.
The PSNR-oriented algorithms are usually based on deep-learning technology.
Comparing with the traditional algorithms, their outputs tend to have sharper edges and higher PSNR performance.
The outputs of GAN-based algorithms are more complicated and challenging for IQA methods.
They do not quite match the quality of detail loss, as they usually contain texture-like noises, or the quality of noise, the texture-like noise is similar to the ground truth in appearance but is not accurate.
An example of GAN-based distortions is shown in \figurename~\ref{fig:GANvis}.
Measuring the similarity of incorrect yet similar features is of great importance to the development of perceptual SR.
The third sub-type includes the outputs of several denoising algorithms.
Similar to image SR, the used denoising algorithms contain both model-based algorithms and deep-learning-based algorithms.
In addition to Gaussian noise, we also include JPEG compression noise removal results.
At last, we include the restoration results of the mixed degradation.
As revealed in \cite{zhang2018learning,qian2019trinity}, performing denoising and SR sequentially will bring new artifacts or different blur effects that barely occur in other IR tasks.

In summary, we have 40 different distortion types and 116 different distortion levels, totally 29k distorted images.
Note that although the number of distortion types is less than some of the existing datasets, we contain a lot of new distortion types and, especially, a large number of IR algorithms' results and GAN results.
This allows the proposed dataset to provide a more objective benchmark for not only IQA methods but also IR methods.

\paragraph{Elo Rating for Mean Opinion Score.}
Given distorted images, the Mean Opinion Score (MOS) is provided for each distorted image.
In literature, there several methodologies used to assess the visual quality of an image \cite{live,tid2013,zhang2018unreasonable,prashnani2018pieapp}.
Early datasets \cite{live} use ``five-gradations rating'' method where images are assigned into five categories directly.
Using this method will result in a huge bias when the users do not have enough experience.
In recent years, datasets usually collect MOS through a large number of pairwise selections using the Swiss rating system \cite{tid2008,tid2013}.
However, as revealed in \cite{prashnani2018pieapp}, the way this pairwise MOS is calculated makes it dependent on specific dataset, it means that the MOS scores of two distorted images can change significantly when they are included in two different datasets.
In order to eliminate this set-dependence effect, Prashnani \textit{et al.} \cite{prashnani2018pieapp} propose to build dataset only based on the probability of pairwise preference.
This method can provide a more accurate propensity probability. 
However, it not only requires a large number of human judgements, but also can not provide the MOS for distortion types, which is important for building benchmarks.
In the proposed dataset, we employ the Elo rating system \cite{elo1978rating} to bring pairwise preference probability and rating system together.
The use of Elo system not only provides reliable human ratings but also reduces the number of required human judgements.

The Elo rating system is a statistic-based rating method and is first proposed for assessing chess player levels.
We assume that the user preference between two images $I_A$ and $I_B$ follows a Logistic distribution parameterized by their Elo Scores \cite{elo20088}.
Given their Elo scores $R_A$ and $R_B$, the expected probability of preference is given by:
\begin{equation}
\label{eq:elo}
    P_{A>B}=\frac{1}{1+10^{(R_B-R_A)/M}}, P_{B>A}=\frac{1}{1+10^{(R_A-R_B)/M}},
\end{equation}
where $P_{A>B}$ indicates the probability that one user would prefer $I_A$ to $I_B$, and $M$ is the parameter of the distribution.
In our dataset we use $M=400$.
Once the user makes a choice, we then update the Elo score for both $I_A$ and $I_B$ use the following rule
\begin{equation}
    R_A'=R_A+K\times(S_A-P_{A>B}), R_B'=R_B+K\times(S_B-P_{B>A}),
\end{equation}
where $K$ is the change step in one judgement and is set to 16.
$S_A$ indicates whether $I_A$ is chosen: $S_A=1$ if $I_A$ wins and $S_A=0$ if $I_A$ fails.
With thousands of human judgements, the Elo scores for each distorted images will converge.
The average of the Elo scores in the last few steps will be assigned as the MOS subjective score.
The averaging operation aims at reducing the randomness of Elo changes.

An example might help understand the Elo system.
Suppose that $R_A=1500$ and $R_b=1600$, then we have $P_{A>B}\approx0.36$ and $P_{B>A}\approx0.64$.
In this situation, if $I_A$ is chosen, the updated Elo score for $I_A$ will be $R_A'=1500 + 16\times(1-0.36)\approx1510$ and the new score for $I_B$ is $R_B'=1600 + 16\times(0-0.64)\approx1594$; if $I_B$ is chosen, the new score will be $R_A'\approx1494$ and $R_B'\approx1605$.
Note that as the expected probabilities for different images being chosen are different, the value changes of the Elo scores will also be different.
This also indicates that when the quality is too different, the winner will not get a lot from winning the bad image.
According to Eq.\eqref{eq:elo}, a score difference of 200 indicates 76\% chance to win, and 400 indicates the chance more than 90\%.
At first, we assign an Elo score of 1400 for each distorted image.
After numerous human judgements (in our dataset, we have 1.13 million human judgements), the Elo score for each image are collected.

Another superiority of employing the Elo system is that our dataset could be dynamic and can be extend in the future.
The Elo system has been widely used to evaluate the relative level of players in electronic games, where the players are constantly changing and the Elo system can provide ratings for new players in a few gameplays.
Recall that one of the chief reasons that ``these IQA methods are facing challenges'' is the invention of GAN and GAN-based IR methods. 
What if other novel image generation technologies are proposed in the future?
Do people need to build a new dataset to include those new algorithms?
With the extendable characteristic of Elo system, one can easily add new distortion types into this dataset and follow the same rating process.
The Elo system will automatically adjust the Elo score for all the distortions without re-rating for the old ones.

\section{Results}
In this section, we conduct a comprehensive study using the proposed PIPAL dataset.
We first build a benchmark for IQA methods.
Through this benchmark, we can answer the question that ``can existing IQA methods objectively evaluate recent IR algorithms?''
We then build a benchmark for some recent SR algorithms to explore the relationship between the development of IQA methods and IR research.
We can get the answer of ``are we getting better IR algorithms by beating benchmarks on these IQA methods?''
At last, we study the characteristics of GAN-based distortion by comparing them with other existing distortion types.
We also improve the performance of IQA networks on GAN-based distortions by introducing anti-aliased pooling layers.

\begin{table}[t]
    \centering
    \footnotesize
    \resizebox{1.0\linewidth}{!}{
    \begin{tabular}{
    p{2.2cm}
    p{2.1cm}<{\centering}
    p{2.1cm}<{\centering}
    % p{2.1cm}<{\centering}
    p{2.1cm}<{\centering}
    p{2.1cm}<{\centering}
    p{2.1cm}<{\centering}
    p{2.1cm}<{\centering}}
    % \begin{tabular}{lccccccc}
    \toprule
        \multirow{2}{*}{\textbf{Method}} &
        \textbf{Traditional} &
        \multirow{2}{*}{\textbf{Denoising}} &
        % \textbf{SR and Denoising} &
        \multirow{2}{*}{\textbf{SR Full}} &
        \textbf{Traditional} &
        \textbf{PSNR.} &
        \textbf{\textit{GAN-based}}
        \\
         &
        \textbf{Distortion} &
         &
        %  &
         &
        \textbf{SR} &
        \textbf{SR} &
        \textbf{\textit{SR}}
        \\
        \hline
        PSNR $\uparrow$         &	0.3589 	&	0.4542 	&	0.4099 	&	0.4782 	&   0.5462 	&	0.2839  \\
        NQM $\uparrow$          &	0.2561 	&	0.5650 	&	0.4742 	&	0.5374 	&	0.6462 	&	0.3410  \\
        UQI $\uparrow$          &	0.3455 	&	0.6246 	&	0.5257 	&	0.6087	&	0.7060 	&	0.3385  \\
        SSIM $\uparrow$         &	0.3910 	&	0.6684 	&	0.5209 	&	0.5856 	&	0.6897 	&	0.3388  \\
        MS-SSIM $\uparrow$      &	0.3967 	&	0.6942 	&	0.5596 	&	0.6527 	&   0.7528 	&	0.3823  \\
        IFC $\uparrow$          &	0.3708 	&	\textbf{0.7440} 	&	0.5651 	&	\textbf{0.7062} 	&	\textbf{0.8244} 	&	0.3217  \\
        VIF $\uparrow$          &	0.4516  &	\textbf{0.7282}     &	0.5917  &	0.6927  &	\textbf{0.7864}  &	0.3857  \\
        VSNR-FR $\uparrow$      &	0.4030  &	0.5938  &	0.5086  &	0.6146  &	0.7076  &	0.3128  \\
        RFSIM $\uparrow$        &	0.3450  &	0.4520  &	0.4232  &	0.4593  &	0.5525  &	0.2951  \\
        GSM $\uparrow$          &	0.5645  &	0.6076  &	0.5361  &	0.6074  &	0.6904  &	0.3523  \\
        SR-SIM $\uparrow$       &	0.6036  &	0.6727  &	0.6094  &	0.6561  &	0.7476  &	0.4631  \\
        FSIM $\uparrow$         &	0.5760  &	0.6882  &	0.5896  &	0.6515  &	0.7381  &	0.4090  \\
        FSIM$_\mathcal{C}$ $\uparrow$              &	0.5724  &	0.6866  &	0.5872  &	0.6509  &	0.7374  &	0.4058  \\
        VSI $\uparrow$                             &	0.4993  &	0.5745  &	0.5475  &	0.6086  &	0.6938  &	0.3706  \\
        MAD $\downarrow$                           &	0.3769  &	0.7005  &	0.5424  &	0.6720  &	0.7575  &	0.3494  \\
        % \hdashline
        LPIPS-Alex $\downarrow$                    &	0.5935  &	0.6688  &	0.5614  &	0.5487  &	0.6782  &	0.4882  \\
        LPIPS-VGG $\downarrow$                     &	0.4087  &	0.7197  &	0.6119  &	0.6077  &	0.7329  &	0.4816  \\
        PieAPP $\downarrow$                        &	\textbf{0.6893}  &	\textbf{0.7435}  &	\textbf{0.7172}  &	\textbf{0.7352}  &	\textbf{0.8097}  &	\textbf{0.5530}  \\
        WaDIQaM $\uparrow$                         &	\textbf{0.6127}  &	0.7157  &	\textbf{0.6621}  &	\textbf{0.6944}  &	0.7628  &	\textbf{0.5343}  \\
        DISTS $\downarrow$                         &	\textbf{0.6213}  &	0.7190  &	\textbf{0.6544}  &	0.6685  &	0.7733  &	\textbf{0.5527}  \\
        \hdashline
        \underline{NIQE} $\downarrow$              &	0.1107  &	-0.0059  &	0.0320  &	0.0599  &	0.1521  &	0.0155  \\
        \underline{Ma \textit{et al.}} $\uparrow$                 &	0.4526  &	0.4963  &	0.3676  &	0.6176  &	0.7124  &	0.0545  \\
        \underline{PI} $\downarrow$                &	0.3631  &	0.3107  &	0.1953  &	0.4833  &	0.5710  &	0.0187  \\
    \toprule
    \end{tabular}}
    \captionsetup{font=small}
    \caption{The SRCC results with respect to different distortion sub-types.
    $\uparrow$ means the higher the better, while  $\downarrow$ means the lower the better. Higher coefficients matche perceptual scores better. The values with top 3 performance are marked in \textbf{blod}
    }
    \label{tab:iqa_sr_results}
\end{table}  % SRCC Table, Big

\begin{figure}[t]
    \centering
    \includegraphics[width=1.0\linewidth]{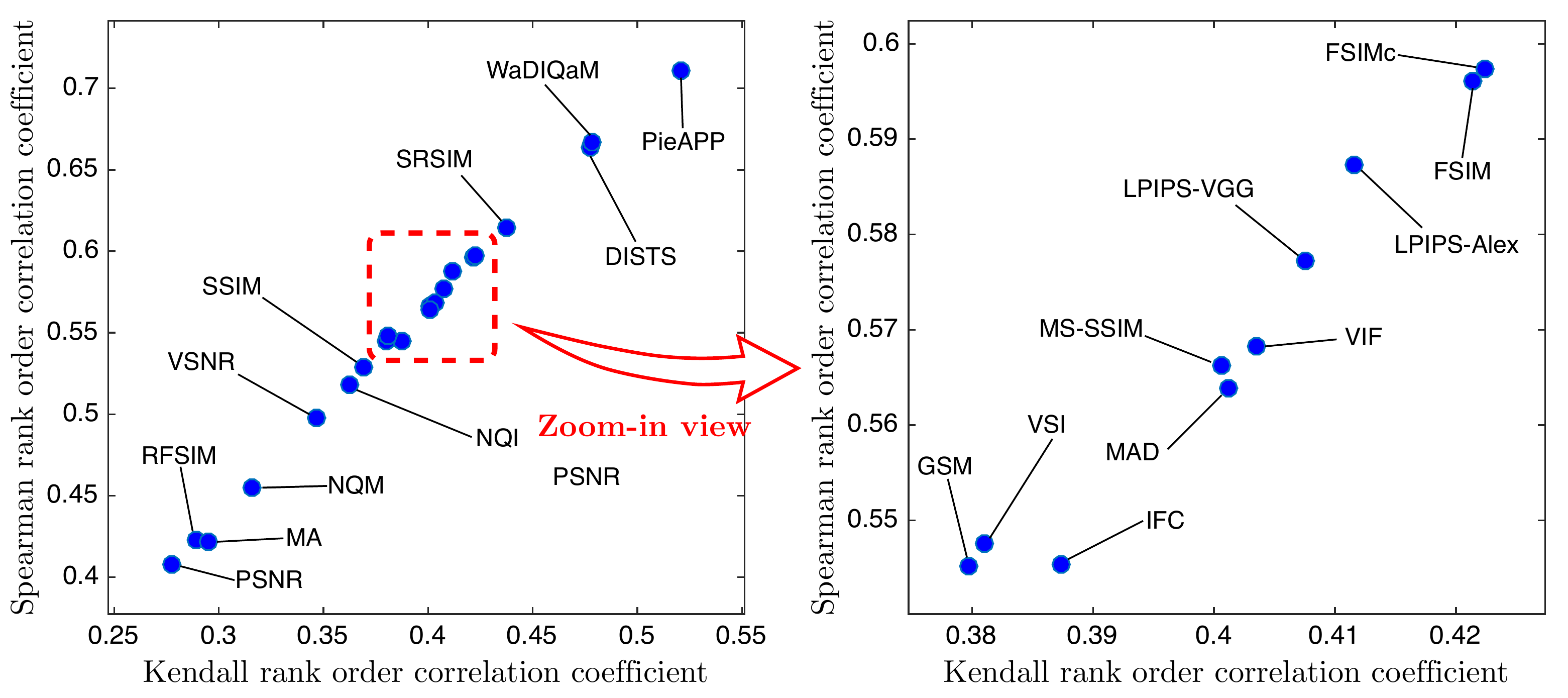}
    \captionsetup{font=small}
    \caption{Quantitative comparison of IQA methods. The right fidgure is the zoom-in view. Higher coefficient matches perceptual score better}
    \label{fig:rcc}
\end{figure}  % Quantitative comparison.

\begin{figure}[t]
    \centering
    \includegraphics[width=1.0\linewidth]{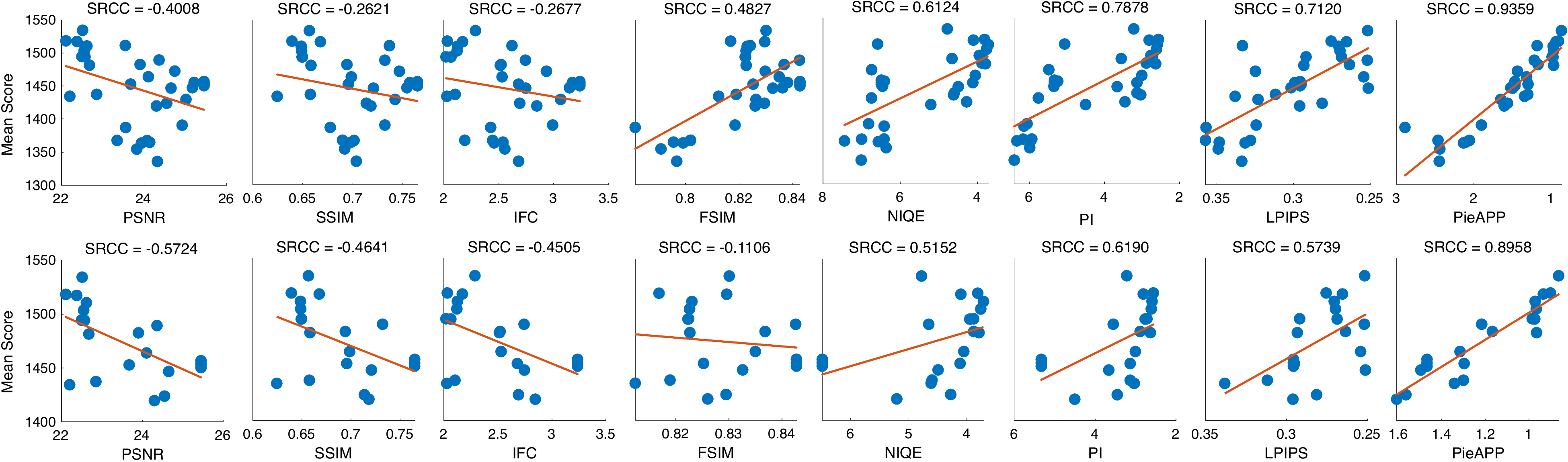}
    \captionsetup{font=small}
    \caption{Analysis of IQA methods in evaluating IR methods. The first row shows the scatter plots of MOS score vs. IQA methods for all SR algorithms. The second row gives scatter plots for GAN-based SR algorithms}
    \label{fig:sr_benchmark}
\end{figure}  % SR Benchmark

\subsection{Evaluation on IQA Methods}
We select a set of commonly-used IQA methods to build the benchmark.
For the FR-IQA methods, we include: PSNR \cite{psnr}, NQM \cite{nqm}, UQI \cite{uqi}, SSIM \cite{ssim}, MS-SSIM \cite{ms-ssim}, IFC \cite{ifc}, VIF \cite{vif}, VSNR-FR \cite{vsnr}, RFSIM \cite{rfsim}, GSM \cite{gsm}, SR-SIM \cite{sr-sim}, FSIM and FSIM$_\mathcal{C}$ \cite{fsim}, SFF \cite{sff}, VSI \cite{vsi}, SCQI \cite{scqi}, LPIPS-Alex and -VGG \cite{zhang2018unreasonable}, PieAPP \cite{prashnani2018pieapp}, WaDIQaM \cite{wadiqam} and DISTS \cite{dists}.
We also include some popular NR-IQA methods: NIQE \cite{niqe}, Ma \cite{ma2017learning}, and PI \cite{blau2018perception}.
All these methods are calculated using the official implementation released by the authors.
As in many previous works \cite{live}, we evaluate IQA methods mainly using Spearman rank order correlation coefficients (SRCC) and Kendall rank order correlation coefficients (KRCC) \cite{kendall1977advanced}.
These two indexes evaluate the monotonicity of methods: whether the scores of high-quality images are higher (or lower) than low-quality images.
We first evaluate the IQA methods using all types of distortions in PIPAL dataset.
A clear exhibition for both SRCC and KRCC rank coefficients is shown in \figurename~\ref{fig:rcc}.
The first conclusion is that even the best IQA method (i.e., PieAPP) provides SRCC with only about 0.71, which is much lower than its performance in TID2013 dataset (about 0.90).
This indicates that the proposed PIPAL dataset is challenging for existing IQA methods and there is a large room for future improvement.
Moreover, a high overall correlation performance does not necessarily indicate the high performance on each sub-type of distortions.
As the focus of this paper, we want to analyze the performance of IQA using IR results, especially the outputs of GAN-based algorithms.
Specifically, we take SR sub-type as an example and show the performance of IQA methods in evaluating SR algorithms.
In \tablename~\ref{tab:iqa_sr_results}, we show the SRCC results with respect to different distortion sub-types, including traditional distortions, denoising outputs, all SR outputs, and the outputs of traditional SR, PSNR-oriented SR and GAN-based SR algorithms.
Analysis of \tablename~\ref{tab:iqa_sr_results} leads to the following conclusions.
First, although performing well in evaluating traditional and PSNR-oriented SR algorithms, almost all IQA methods suffer from severe performance drop when evaluating GAN-based algorithms.
This confirms the conclusion of Blua \textit{et al.} \cite{blau2018perception} that less distortion (e.g., higher PSNR values) may be related to lower perceptual performance for GAN-based IR algorithms.
Second, despite of the severe performance drop, several IQA methods still outperform the others on GAN-based algorithms.
Coincidentally, they are all recent works and based on deep networks.

We then present the analysis of IQA methods as IR evaluation metrics.
In \figurename~\ref{fig:sr_benchmark}, we show the scatter plots of subjective scores vs. the average values of some commonly-used image quality metrics for 23 SR algorithms.
Among them, PSNR and SSIM are the most common measures, IFC is suggested by Yang \textit{et al.} \cite{yang2014single}, NIQE and PI are suggested in recent works \cite{blau2018perception,zhang2019ranksrgan} for their good performance on GAN-based SR algorithms.
LPIPS \cite{zhang2018unreasonable} and PieAPP \cite{prashnani2018pieapp} are selected according to our benchmark.
As can be seen that, although widely used, PSNR, SSIM and IFC are anti-correlated with the subjective scores, thus are inappropriate for evaluating GAN-based algorithms.
It is worth noting that IFC shows good performance on denoising, traditional SR and PSNR-oriented SR according to \tablename~\ref{tab:iqa_sr_results}, but drops severely on GAN-based distortions.
NIQE and PI show moderate performance on evaluating IR algorithms, and LPIPS and PieAPP are the most correlated.
Note that different from the work of Blau \textit{et al} \cite{blau2018perception} where they evaluate the perceptual quality only based on whether the image looks real, we collect subjective scores based on the perceptual similarity with the ground truth.
Therefore, in evaluating the performance of the IR algorithms from the perspective of reconstruction, the suggestions given by our work are more appropriate.
%
% These results have a similar trend as the results presented in \figurename~\ref{fig:scatter}.

\begin{table}[t]
\begin{center}
    \footnotesize
    {\resizebox{\linewidth}{!}{
  \begin{tabular}{
  p{2.6cm}
  p{1.4cm}%<{\centering}
  p{1.4cm}%<{\centering}
  p{1.4cm}%<{\centering}
  p{1.4cm}%<{\centering}
  p{1.4cm}%<{\centering}
  p{1.4cm}%<{\centering}
  p{1.4cm}%<{\centering}
%   p{1.4cm}%<{\centering}
  p{1.4cm}%<{\centering}
  }
        \toprule
        % \hline
        Method &
        Year &
        PSNR $\uparrow$ &
        ~SSIM $\uparrow$ &
        \underline{Ma} $\uparrow$ &
        ~\underline{NIQE} $\downarrow$ &
        ~~~~\underline{PI} $\downarrow$ &
        LPIPS $\downarrow$ &
        % \textbf{PieAPP} $\downarrow$ &
        ~~MOS $\uparrow$ \\
        \specialrule{0em}{1pt}{1pt}
        \hline
        \specialrule{0em}{1pt}{1pt}
        YY \cite{yy2013}          & 2013 & 23.35$^8$    & 0.6897$^7$    & 4.5486$^{10}$ & 6.4174$^8$    & 5.9344$^7$    & 0.3574$^{12}$ & 1367.71$^8$ \\
        TSG \cite{tsg2013}         & 2013 & 23.55$^7$    & 0.6775$^8$    & 4.1298$^{12}$ & 6.4163$^7$    & 6.1433$^{10}$ & 0.3570$^{11}$ & 1387.24$^7$ \\
        A+ \cite{a+2014}         & 2014 & 23.82$^6$    & 0.6919$^6$    & 4.3852$^{11}$ & 6.3645$^5$    & 5.9897$^9$    & 0.3491$^{10}$ & 1354.52$^{12}$ \\
        SRCNN \cite{srcnn2014}      & 2014 & 23.93$^5$    & 0.6966$^5$    & 4.6094$^9$    & 6.5657$^{10}$ & 5.9781$^8$    & 0.3316$^8$    & 1363.68$^{11}$ \\
        FSRCNN \cite{fsrcnn2016}     & 2016 & 24.07$^4$    & 0.7013$^3$    & 4.6686$^8$    & 6.9985$^{11}$ & 6.1649$^{11}$ & 0.3281$^7$    & 1367.49$^9$ \\
        VDSR \cite{vdsr2016}      & 2016 & 24.13$^3$    & 0.6984$^4$    & 4.7799$^7$    & 7.4436$^{12}$ & 6.3319$^{12}$ & 0.3484$^9$    & 1364.90$^{10}$ \\
        EDSR \cite{edsr2017}      & 2017 & \textbf{25.17}$^2$    & \textbf{0.7541}$^2$    & 5.7634$^6$    & 6.4560$^9$    & 5.3463$^6$    & 0.3016$^6$    & 1447.44$^6$ \\
        SRGAN \cite{srgan2017}     & 2017 & 22.57$^{10}$ & 0.6494$^{11}$ & 8.4215$^3$    & 3.9527$^3$    & 2.7656$^3$    & \textbf{0.2687}$^2$    & 1494.14$^3$ \\
        RCAN \cite{rcan2018}     & 2018 & \textbf{25.21}$^1$    & \textbf{0.7569}$^1$    & 5.9260$^5$    & 6.4121$^6$    & 5.2430$^5$    & 0.2992$^5$    & 1455.31$^5$ \\
        BOE \cite{boe2018}      & 2018 & 22.68$^9$    & 0.6582$^9$    & \textbf{8.5209}$^2$    & \textbf{3.7945}$^1$    & \textbf{2.6368}$^2$    & 0.2933$^4$    & 1481.51$^4$ \\
        ESRGAN \cite{wang2018esrgan} & 2018 & 22.51$^{11}$ & 0.6566$^{10}$ & 8.3424$^4$    & 4.7821$^4$    & 3.2198$^4$    & \textbf{0.2517}$^1$    & \textbf{1534.25}$^1$ \\
        RankSRGAN \cite{zhang2019ranksrgan} & 2019 & 22.11$^{12}$ & 0.6392$^{12}$ & \textbf{8.6882}$^1$    & \textbf{3.8155}$^2$    & \textbf{2.5636}$^1$    & 0.2755$^3$    & \textbf{1518.29}$^2$ \\
        \toprule
        % \hline
    \end{tabular}
    }}
    \captionsetup{font=small}
    \caption{The $\times4$ SR results. The years of publication are also provided. The \textbf{bolded} values are the top 2 values and the superscripts indicate the ranking
    }
    \label{tab:sr_evaluation}
\end{center}
\end{table}  % SR Evaluation

\subsection{Evaluation on IR Methods}
\label{sec:evaluation_IR}
One of the most important applications of IQA technology is to evaluate IR algorithms.
IQA methods have been the chief reason for the progress in the IR field as a means of comparing the performance.
However, evaluating IR methods only with specific IQA methods also narrows the focus of IR research and converts it to competitions only on the quantitative numbers (e.g., PSNR competitions \cite{timofte2017ntire,cai2019ntire} and PI competition \cite{blau20182018}).
As stated above, existing IQA methods may be inadequate in evaluating IR algorithms.
We wonder that with the focus on beating benchmarks on the flawed IQA methods, are we getting better IR algorithms?
To answer this question, we take SR task as a representative and select 12 SR algorithms to build a benchmark.
These are all representative algorithms and selected from 2013 to the present.
The results are shown in \tablename~\ref{tab:sr_evaluation}.
We present more algorithms in the Supplementary Material.
One can observe that before 2017 (when GAN was applied to SR) the PSNR performance improves continuously.
Especially, the deep-learning-based algorithms improve PSNR by about 1.4dB.
These efforts do improve the subjective performance -- the average MOS value increases by about 90 in 4 years.
After SRGAN was proposed, the PSNR decreased by about 2.6dB compared to the state-of-the-art PSNR performance at that time (EDSR), but the MOS value increased by about 50 suddenly.
In contrast, RCAN was proposed to defeat EDSR in terms of PSNR.
Its PSNR performance is a little higher than EDSR but its MOS score is even lower than EDSR.
When noticing that the mainstream metrics (PSNR and SSIM) had conflicted with the subjective performance, PI was proposed to evaluate perceptual SR algorithms \cite{blau2018perception}.
After that, ESRGAN and RankSRGAN have been continuously improving PI performance.
Among them, the latest RankSRGAN has achieved the current state-of-the-art in terms of PI and NIQE performance.
However, ESRGAN has the highest subjective performance, but has no advantage in terms of PI and NIQE comparing with RankSRGAN.
Efforts on improving the PI value show limited effects and have failed to continuously improve MOS scores after ESRGAN.
These observations inspire us in two aspects.
First, none of existing IQA methods is always effective in evaluation.
With the development of IR technology, new IQA methods need to be proposed accordingly.
Second, excessively optimizing performance on a specific IQA may cause a decrease in perceptual quality.

\begin{figure}[t]
    \centering
    \includegraphics[width=1.0\linewidth]{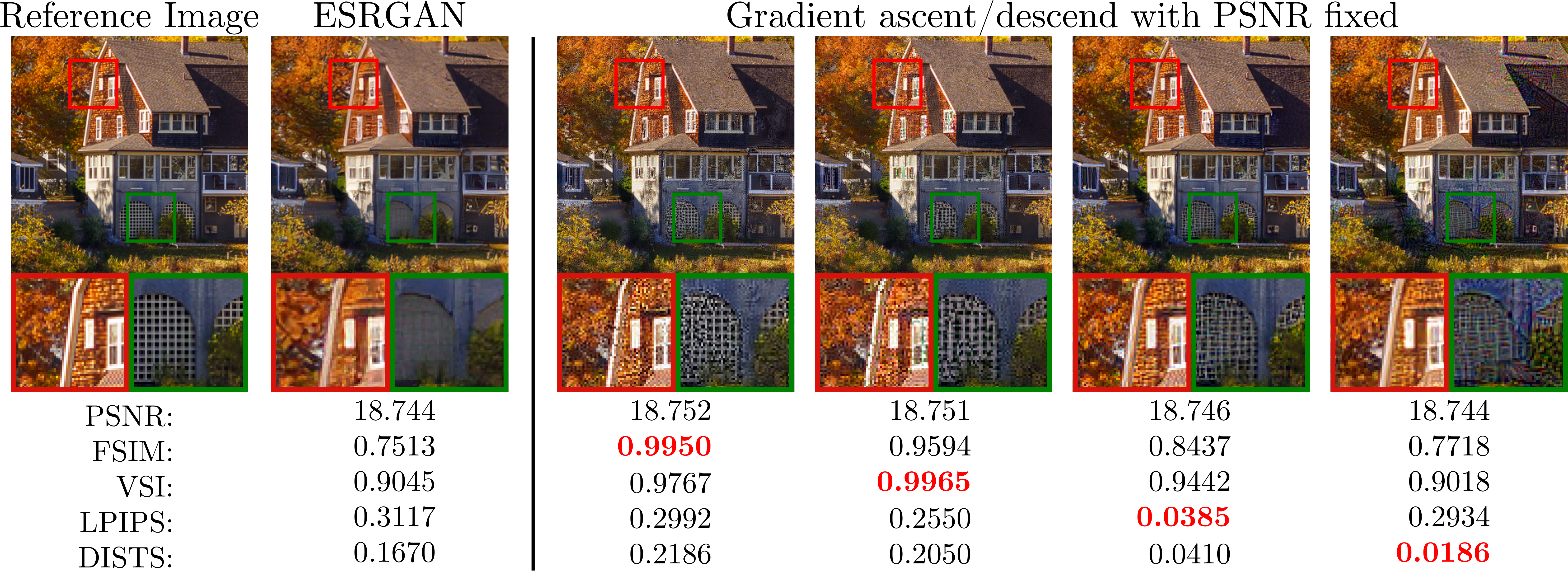}
    \captionsetup{font=small}
    \caption{Best-case images with respect to different IQA methods, with identical PSNR. These are computed by gradient ascent/descent optimization on certain IQA methods}
    \label{fig:iqa_optim}
\end{figure}

We conduct experiments to explore this possibility by performing gradient ascend/descend of certain IQA methods.
According to Blau \textit{et al.} \cite{blau2018perception}, distortion and perceptual quality are at odds with each other.
In order to simulate the situation where there is a perception-distortion trade-off, we constrain the PSNR value to be equal to that of the initial distorted image during optimization.
We use the output of ESRGAN as the initial image, and the results are shown in \figurename~\ref{fig:iqa_optim}.
We can see that some images show superior numerical performance when evaluated using certain IQA methods, but may not be dominant in other metrics.
Their best-cases also show different visual effects.
Even for some IQA methods (LPIPS and DISTS) with good performance on GAN-based distortion, their best-cases still contain serious artifacts.
This indicates that evaluating and developing new IQA methods plays an important role in future research.

  % IQA Optim

\subsection{Discussion of GAN-based Distortion}
\label{sec:GANDistort}
Recall that LPIPS, PieAPP and DISTS perform relatively better in evaluating GAN-based distortion. 
The effectiveness of these methods may be attributed to the following reasons.
Compared with other IQA methods, deep-learning-based IQA methods can extract image features more effectively.
For traditional distortion types, such as blur, compression and noise, the distorted images usually disobey the prior distribution of natural images.
Early IQA methods can assess these images by measuring low-level statistic image features such as image gradient and structural information.
These strategies are also effective for the outputs of traditional and PSNR-oriented algorithms.
However, most of these strategies fail in the case of GAN-based distortion, as the way that GAN-based distortions differs from the reference images is less apparent.
They may have similar image statistic features with the reference image.
In this case, deep networks are able to capture these unapparent features and distinguish such distortions to some extent.

\begin{figure}[t]
    \centering
    \setlength{\tabcolsep}{0.0mm}
    \resizebox{1.0\linewidth}{!}{
    \begin{tabular}{cccc}
    \includegraphics[width=0.25\linewidth]{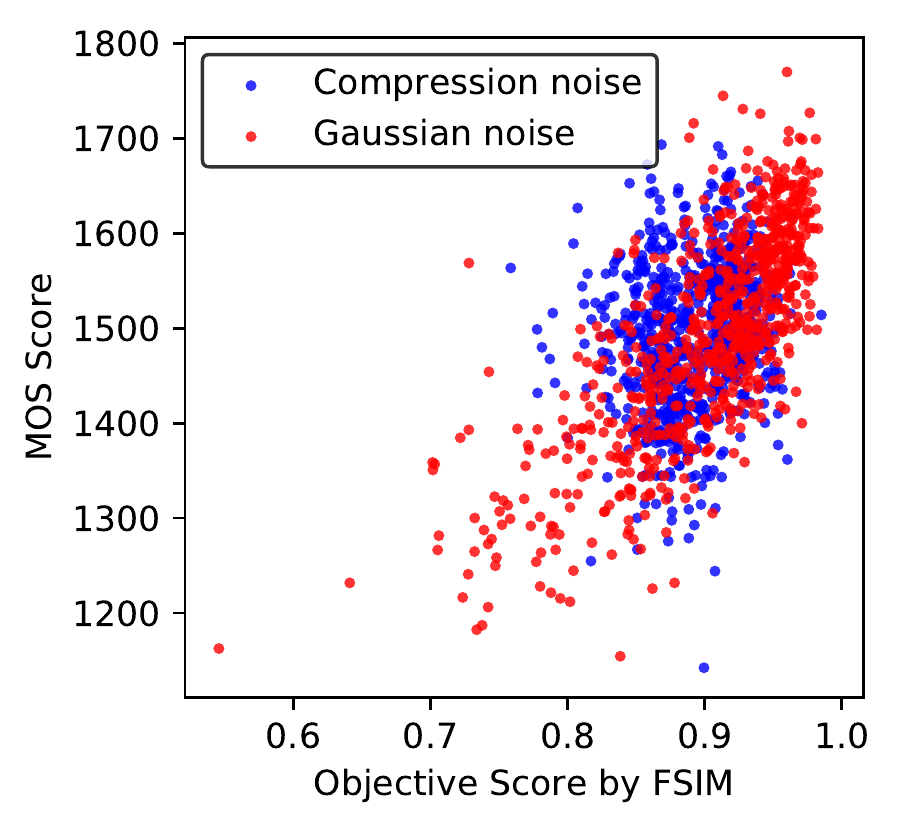} &
    \includegraphics[width=0.25\linewidth]{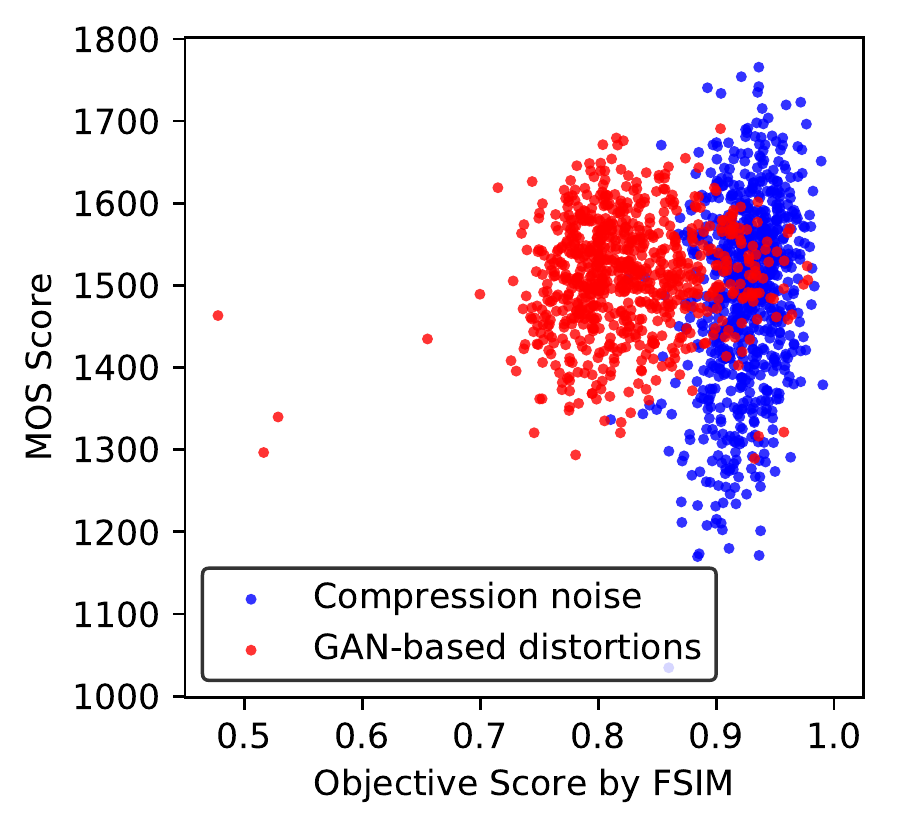} &
    \includegraphics[width=0.25\linewidth]{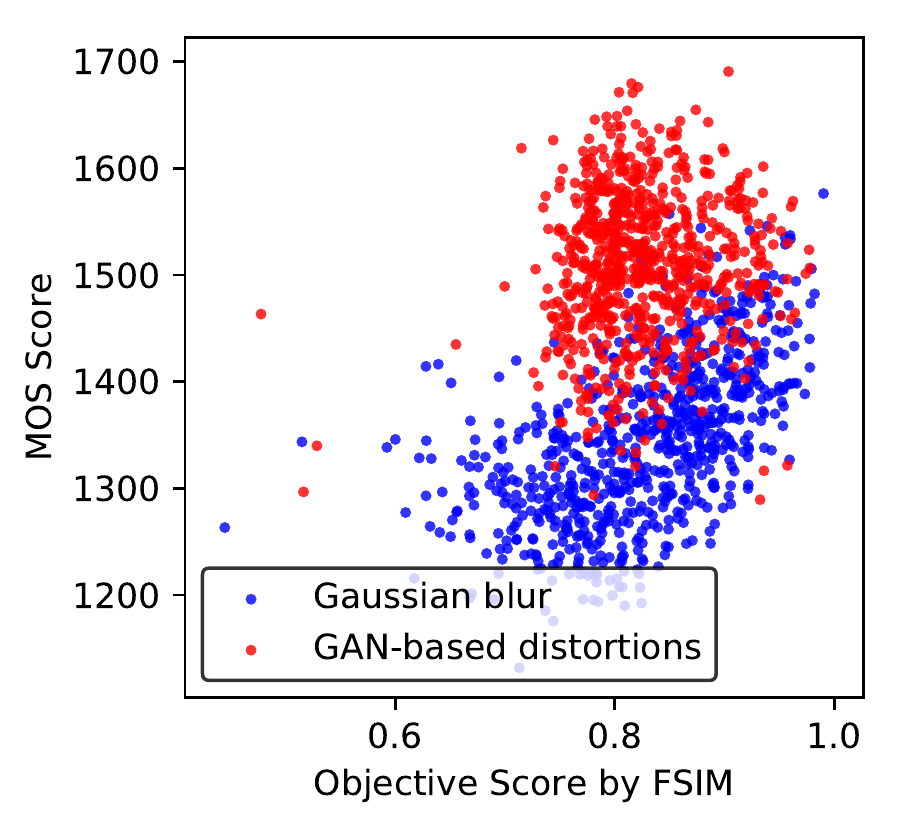} &
    \includegraphics[width=0.25\linewidth]{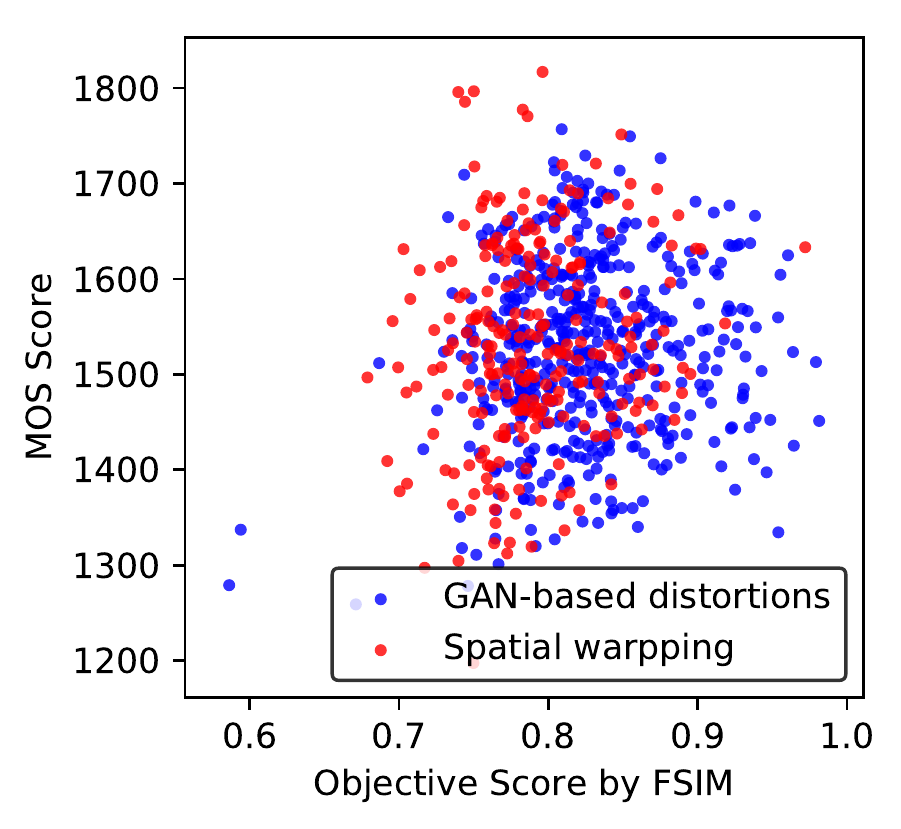} \\
    \scriptsize (a) &
    \scriptsize (b) &
    \scriptsize (c) &
    \scriptsize (d) \\
    \end{tabular}
    }
    \captionsetup{font=small}
    \caption{Examples of scatter plots for pairs of distortion types. For distortion types that are easy to measure, samples are well clustered along the fitted curve. For others that are difficult for IQA method, the samples will not be well clustered. The samples of distortion types which have similar behavior will overlap with each other}
    \label{fig:gan_warp}
\end{figure}  % GAN distortion vs. Warping

In order to explore the characteristics of GAN-based distortion, we compare them with some well-studied distortions.
As stated in \cite{tid2013}, for a good IQA method, the subjective scores in the scatter plot should increase coincide with objective values, and the samples are well clustered along the fitted curve.
In the case of two distortion types, if the IQA method behaves similarly for both of them, their samples on the scatter plot will also be well clustered and overlaped.
For example, the additive Gaussian noise and lossy compression are well studied distortion types for most IQA methods.
When calculating the objective values using FSIM$_\mathcal{C}$, samples of both distortions are clustered, as shown in \figurename~\ref{fig:gan_warp} (a).
This indicates that FSIM$_\mathcal{C}$ can adequately characterize the visual quality of an image that is damaged due to these two types of distortion.
Then we study GAN-based distortion by comparing it with some existing distortion types using FSIM$_\mathcal{C}$: \figurename~\ref{fig:gan_warp} (b) shows the result of GAN-based distortion and additive Gaussian noise, and \figurename~\ref{fig:gan_warp} (c) shows the result of GAN-based distortion and Gaussian blur.
It can be seen that the samples of Gaussian noise and Gaussian blur barely intersect with GAN-based samples.
FSIM$_\mathcal{C}$ largely underestimates the visual quality of GAN-based distortion.
In \figurename~\ref{fig:gan_warp} (d), we show the result of GAN-based distortion and spatial warping distortion.
As can be seen, these two distortion types behave unexpectedly similar.
FSIM$_\mathcal{C}$ cannot handle them and presents the same random and diffused state.
The quantitative results also verify this phenomenon.
For spatial warping distortion type, the SRCC of FSIM$_\mathcal{C}$ is 0.31, and it is close to the performance of GAN-based distortion, which is 0.41.
Thus we argue that the spatial warping distortion and GAN-based distortion pose similar challenges to FSIM$_\mathcal{C}$.

As revealed in experimental psychology \cite{kolers1962intensity,kahneman1968method}, the mutual interference between visual information may cause the Visual Masking effects.
According to this theory, some key reasons that IQA methods tend to underestimate both GAN-based distortion and spatial warping distortion are as follows.
Firstly, for the edges with strong intensity change, the human visual system (HVS) is sensitive to the contour and shape, but not sensitive to the error and misalignment of the edges.
Secondly, the ability of HVS to distinguish texture decreases in the region with dense textures.
When the extracted features of the textures are similar, the HVS will ignore part of the subtle differences and misalignment of textures.
However, most of the traditional and deep-learning-based IQA methods require good alignment for the inputs.
This partially causes the drop of performance of these IQA methods on GAN-based distortion.

This finding provides us an insight that if we explicitly consider the spatial misalignment, we may improve the performance of IQA methods on GAN-based distortion.
We explore this possibility by introducing anti-aliasing pooling layer to IQA networks.
IQA networks extract features by cascaded convolution operations.
If we want the IQA networks to be robust to small misalignment, the extracted features should at least be invariant to this misalignment/shift.
CNN should have been shift-invariant, as the standard convolution operations are shift-invariant.
However, IQA networks are usually not shift-invariant, as some commonly used downsampling layers, such as max/average pooling and strided-convolution ignore the sampling theorem \cite{zhang2019making}.
These operations are employed in VGG \cite{simonyan2014very} and Alex \cite{krizhevsky2012imagenet} networks, which are popular backbone architectures for feature extraction (e.g., in LPIPS and DISTS).
As suggested by Zhang \cite{zhang2019making}, one can fix this by introducing anti-aliasing pooling.
We conduct this experiment based on LPIPS-Alex and introduce l2 pooling \cite{dists} and BlurPool \cite{zhang2019making} layers to replace its max pooling layers. 
l2 pooling fixes this problem by low-pass filtering before downsampling, and BlurPool further improves the performance by low-pass filtering between dense max operation and subsampling.
The results are shown in \tablename~\ref{tab:iqa_pool}.
We observe increased correlation in both PIPAL full set and GAN-based distortion subset.
The results demonstrate the effectiveness of improving anti-aliasing pooling and indicate that the lack of robustness to small misalignment is one of the reasons for the decline in the performance of GAN-based distortion.

\begin{table}[t]
    \centering
    \resizebox{1.0\linewidth}{!}{
    \begin{tabular}{
        p{3.0cm}|
        p{1.7cm}<{\centering}
        p{1.7cm}<{\centering}|
        p{1.7cm}<{\centering}
        p{1.7cm}<{\centering}|
        p{1.7cm}<{\centering}
        p{1.7cm}<{\centering}}
         \multirow{2}{*}{Test Set} & \multicolumn{2}{c|}{LPIPS baseline} & \multicolumn{2}{c|}{LPIPS + l2 pooling} & \multicolumn{2}{c}{LPIPS + BlurPool} \\
             & SRCC & KRCC & SRCC & KRCC & SRCC & KRCC \\
            \hline
            PIPAL & 0.5604 & 0.3910 & 0.5816 & 0.4080 & \textbf{0.5918} & \textbf{0.4160}  \\
            PIPAL GAN distort. & 0.4862 & 0.3339 & 0.4942 & 0.3394 & \textbf{0.5135} & \textbf{0.3549}  \\
    \end{tabular}
    }
    \captionsetup{font=small}
    \caption{The SRCC and KRCC performance of LPIPS with different pooling layers. The anti-aliasing pooling layers (l2 pooling \cite{dists} and BlurPool \cite{zhang2019making}) improve the performance both on the PIPAL full set and GAN-based distortion subset}
    \label{tab:iqa_pool}
\end{table}

\section{Conclusion}
In this paper, we construct a novel IQA dataset, namely PIPAL and establish benchmarks for both IQA methods and IR algorithms.
Our results indicate that existing IQA methods face challenges in evaluating perceptual IR algorithms, especially GAN-based algorithms.
We also shed light on improving IQA networks by introducing anti-aliasing pooling layers.
Experiments demonstrate the effectiveness of the proposed strategy.

\subsubsection{Acknowledgement.}
This work is partially supported by SenseTime Group Limited, the National Natural Science Foundation of China (61906184), Science and Technology Service Network Initiative of Chinese Academy of Sciences (KFJ-STS-QYZX-092), Shenzhen Basic Research Program (JSGG20180507182100698, CXB201104220032A), the Joint Lab of CAS-HK, Shenzhen Institute of Artificial Intelligence and Robotics for Society.

\bibliographystyle{splncs04}
\bibliography{egbib.bib}

\begin{thebibliography}{10}
\providecommand{\url}[1]{\texttt{#1}}
\providecommand{\urlprefix}{URL }
\providecommand{\doi}[1]{https://doi.org/#1}

\bibitem{div2k}
Agustsson, E., Timofte, R.: Ntire 2017 challenge on single image
  super-resolution: Dataset and study. In: Proceedings of the IEEE Conference
  on Computer Vision and Pattern Recognition Workshops. pp. 126--135 (2017)

\bibitem{scqi}
Bae, S.H., Kim, M.: A novel image quality assessment with globally and locally
  consilient visual quality perception. IEEE Transactions on Image Processing
  \textbf{25}(5),  2392--2406 (2016)

\bibitem{blau20182018}
Blau, Y., Mechrez, R., Timofte, R., Michaeli, T., Zelnik-Manor, L.: The 2018
  pirm challenge on perceptual image super-resolution. In: European Conference
  on Computer Vision. pp. 334--355. Springer (2018)

\bibitem{blau2018perception}
Blau, Y., Michaeli, T.: The perception-distortion tradeoff. In: Proceedings of
  the IEEE Conference on Computer Vision and Pattern Recognition. pp.
  6228--6237 (2018)

\bibitem{wadiqam}
Bosse, S., Maniry, D., M{\"u}ller, K.R., Wiegand, T., Samek, W.: Deep neural
  networks for no-reference and full-reference image quality assessment. IEEE
  Transactions on Image Processing  \textbf{27}(1),  206--219 (2017)

\bibitem{cai2019ntire}
Cai, J., Gu, S., Timofte, R., Zhang, L., Liu, X., Ding, Y., He, D., Li, C., Fu,
  Y., Wen, S., et~al.: Ntire 2019 challenge on real image super-resolution:
  Methods and results. In: 2019 IEEE/CVF Conference on Computer Vision and
  Pattern Recognition Workshops (CVPRW). pp. 2211--2223. IEEE (2019)

\bibitem{vsnr}
Chandler, D.M., Hemami, S.S.: Vsnr: A wavelet-based visual signal-to-noise
  ratio for natural images. IEEE transactions on image processing
  \textbf{16}(9),  2284--2298 (2007)

\bibitem{sff}
Chang, H.W., Yang, H., Gan, Y., Wang, M.H.: Sparse feature fidelity for
  perceptual image quality assessment. IEEE Transactions on Image Processing
  \textbf{22}(10),  4007--4018 (2013)

\bibitem{bm3d}
Dabov, K., Foi, A., Katkovnik, V., Egiazarian, K.: Image denoising by sparse
  3-d transform-domain collaborative filtering. IEEE Transactions on image
  processing  \textbf{16}(8),  2080--2095 (2007)

\bibitem{nqm}
Damera-Venkata, N., Kite, T.D., Geisler, W.S., Evans, B.L., Bovik, A.C.: Image
  quality assessment based on a degradation model. IEEE transactions on image
  processing  \textbf{9}(4),  636--650 (2000)

\bibitem{dists}
Ding, K., Ma, K., Wang, S., Simoncelli, E.P.: Image quality assessment:
  Unifying structure and texture similarity. arXiv preprint arXiv:2004.07728
  (2020)

\bibitem{srcnn2014}
Dong, C., Loy, C.C., He, K., Tang, X.: Image super-resolution using deep
  convolutional networks. IEEE transactions on pattern analysis and machine
  intelligence  \textbf{38}(2),  295--307 (2015)

\bibitem{fsrcnn2016}
Dong, C., Loy, C.C., Tang, X.: Accelerating the super-resolution convolutional
  neural network. In: European conference on computer vision. pp. 391--407.
  Springer (2016)

\bibitem{elo1978rating}
Elo, A.E.: The rating of chessplayers, past and present. Arco Pub. (1978)

\bibitem{elo20088}
Elo, A.E.: Logistic probability as a rating basis. The Rating of Chessplayers,
  Past\&Present. Bronx NY  \textbf{10453} (2008)

\bibitem{feng2019suppressing}
Feng, R., Gu, J., Qiao, Y., Dong, C.: Suppressing model overfitting for image
  super-resolution networks. In: 2019 IEEE/CVF Conference on Computer Vision
  and Pattern Recognition Workshops (CVPRW). pp. 1964--1973. IEEE (2019)

\bibitem{goodfellow2014generative}
Goodfellow, I., Pouget-Abadie, J., Mirza, M., Xu, B., Warde-Farley, D., Ozair,
  S., Courville, A., Bengio, Y.: Generative adversarial nets. In: Advances in
  neural information processing systems. pp. 2672--2680 (2014)

\bibitem{psnr}
Group, V.Q.E., et~al.: Final report from the video quality experts group on the
  validation of objective models of video quality assessment. In: VQEG meeting,
  Ottawa, Canada, March, 2000 (2000)

\bibitem{gu2019blind}
Gu, J., Lu, H., Zuo, W., Dong, C.: Blind super-resolution with iterative kernel
  correction. In: Proceedings of the IEEE Conference on Computer Vision and
  Pattern Recognition. pp. 1604--1613 (2019)

\bibitem{gu2019image}
Gu, J., Shen, Y., Zhou, B.: Image processing using multi-code gan prior. In:
  Proceedings of the IEEE/CVF Conference on Computer Vision and Pattern
  Recognition. pp. 3012--3021 (2020)

\bibitem{jain2009natural}
Jain, V., Seung, S.: Natural image denoising with convolutional networks. In:
  Advances in neural information processing systems. pp. 769--776 (2009)

\bibitem{kahneman1968method}
Kahneman, D.: Method, findings, and theory in studies of visual masking.
  Psychological Bulletin  \textbf{70}(6p1), ~404 (1968)

\bibitem{kendall1977advanced}
Kendall, M., Stuart, A.: The advanced theory of statistics; charles griffin \&
  co. Ltd.(London)  \textbf{83},  62013 (1977)

\bibitem{vdsr2016}
Kim, J., Kwon~Lee, J., Mu~Lee, K.: Accurate image super-resolution using very
  deep convolutional networks. In: Proceedings of the IEEE conference on
  computer vision and pattern recognition. pp. 1646--1654 (2016)

\bibitem{kolers1962intensity}
Kolers, P.A.: Intensity and contour effects in visual masking. Vision Research
  \textbf{2}(9-10),  277--IN4 (1962)

\bibitem{krizhevsky2012imagenet}
Krizhevsky, A., Sutskever, I., Hinton, G.E.: Imagenet classification with deep
  convolutional neural networks. In: Advances in neural information processing
  systems. pp. 1097--1105 (2012)

\bibitem{csiq}
Larson, E.C., Chandler, D.M.: Most apparent distortion: full-reference image
  quality assessment and the role of strategy. Journal of Electronic Imaging
  \textbf{19}(1),  011006 (2010)

\bibitem{srgan2017}
Ledig, C., Theis, L., Husz{\'a}r, F., Caballero, J., Cunningham, A., Acosta,
  A., Aitken, A., Tejani, A., Totz, J., Wang, Z., et~al.: Photo-realistic
  single image super-resolution using a generative adversarial network. In:
  Proceedings of the IEEE conference on computer vision and pattern
  recognition. pp. 4681--4690 (2017)

\bibitem{edsr2017}
Lim, B., Son, S., Kim, H., Nah, S., Mu~Lee, K.: Enhanced deep residual networks
  for single image super-resolution. In: Proceedings of the IEEE conference on
  computer vision and pattern recognition workshops. pp. 136--144 (2017)

\bibitem{gsm}
Liu, A., Lin, W., Narwaria, M.: Image quality assessment based on gradient
  similarity. IEEE Transactions on Image Processing  \textbf{21}(4),
  1500--1512 (2011)

\bibitem{ma2017learning}
Ma, C., Yang, C.Y., Yang, X., Yang, M.H.: Learning a no-reference quality
  metric for single-image super-resolution. Computer Vision and Image
  Understanding  \textbf{158},  1--16 (2017)

\bibitem{boe2018}
Michelini, P.N., Zhu, D., Liu, H.: Multi--scale recursive and
  perception--distortion controllable image super--resolution. In: European
  Conference on Computer Vision. pp. 3--19. Springer (2018)

\bibitem{brisque}
Mittal, A., Moorthy, A.K., Bovik, A.C.: No-reference image quality assessment
  in the spatial domain. IEEE Transactions on image processing
  \textbf{21}(12),  4695--4708 (2012)

\bibitem{niqe}
Mittal, A., Soundararajan, R., Bovik, A.C.: Making a “completely blind”
  image quality analyzer. IEEE Signal Processing Letters  \textbf{20}(3),
  209--212 (2012)

\bibitem{tid2013}
Ponomarenko, N., Jin, L., Ieremeiev, O., Lukin, V., Egiazarian, K., Astola, J.,
  Vozel, B., Chehdi, K., Carli, M., Battisti, F., et~al.: Image database
  tid2013: Peculiarities, results and perspectives. Signal Processing: Image
  Communication  \textbf{30},  57--77 (2015)

\bibitem{tid2008}
Ponomarenko, N., Lukin, V., Zelensky, A., Egiazarian, K., Carli, M., Battisti,
  F.: Tid2008-a database for evaluation of full-reference visual quality
  assessment metrics. Advances of Modern Radioelectronics  \textbf{10}(4),
  30--45 (2009)

\bibitem{prashnani2018pieapp}
Prashnani, E., Cai, H., Mostofi, Y., Sen, P.: Pieapp: Perceptual image-error
  assessment through pairwise preference. In: Proceedings of the IEEE
  Conference on Computer Vision and Pattern Recognition. pp. 1808--1817 (2018)

\bibitem{qian2019trinity}
Qian, G., Gu, J., Ren, J.S., Dong, C., Zhao, F., Lin, J.: Trinity of pixel
  enhancement: a joint solution for demosaicking, denoising and
  super-resolution. arXiv preprint arXiv:1905.02538  (2019)

\bibitem{enhancenet2017}
Sajjadi, M.S., Scholkopf, B., Hirsch, M.: Enhancenet: Single image
  super-resolution through automated texture synthesis. In: Proceedings of the
  IEEE International Conference on Computer Vision. pp. 4491--4500 (2017)

\bibitem{vif}
Sheikh, H.R., Bovik, A.C.: Image information and visual quality. IEEE
  Transactions on image processing  \textbf{15}(2),  430--444 (2006)

\bibitem{ifc}
Sheikh, H.R., Bovik, A.C., De~Veciana, G.: An information fidelity criterion
  for image quality assessment using natural scene statistics. IEEE
  Transactions on image processing  \textbf{14}(12),  2117--2128 (2005)

\bibitem{live}
Sheikh, H.R., Sabir, M.F., Bovik, A.C.: A statistical evaluation of recent full
  reference image quality assessment algorithms. IEEE Transactions on image
  processing  \textbf{15}(11),  3440--3451 (2006)

\bibitem{simonyan2014very}
Simonyan, K., Zisserman, A.: Very deep convolutional networks for large-scale
  image recognition. arXiv preprint arXiv:1409.1556  (2014)

\bibitem{timofte2017ntire}
Timofte, R., Agustsson, E., Van~Gool, L., Yang, M.H., Zhang, L.: Ntire 2017
  challenge on single image super-resolution: Methods and results. In:
  Proceedings of the IEEE Conference on Computer Vision and Pattern Recognition
  Workshops. pp. 114--125 (2017)

\bibitem{tsg2013}
Timofte, R., De~Smet, V., Van~Gool, L.: Anchored neighborhood regression for
  fast example-based super-resolution. In: Proceedings of the IEEE
  international conference on computer vision. pp. 1920--1927 (2013)

\bibitem{a+2014}
Timofte, R., De~Smet, V., Van~Gool, L.: A+: Adjusted anchored neighborhood
  regression for fast super-resolution. In: Asian conference on computer
  vision. pp. 111--126. Springer (2014)

\bibitem{wang2018esrgan}
Wang, X., Yu, K., Wu, S., Gu, J., Liu, Y., Dong, C., Qiao, Y., Loy, C.C.:
  Esrgan: Enhanced super-resolution generative adversarial networks. In:
  European Conference on Computer Vision. pp. 63--79. Springer (2018)

\bibitem{uqi}
Wang, Z., Bovik, A.C.: A universal image quality index. IEEE signal processing
  letters  \textbf{9}(3),  81--84 (2002)

\bibitem{ssim}
Wang, Z., Bovik, A.C., Sheikh, H.R., Simoncelli, E.P., et~al.: Image quality
  assessment: from error visibility to structural similarity. IEEE transactions
  on image processing  \textbf{13}(4),  600--612 (2004)

\bibitem{ms-ssim}
Wang, Z., Simoncelli, E.P., Bovik, A.C.: Multiscale structural similarity for
  image quality assessment. In: The Thrity-Seventh Asilomar Conference on
  Signals, Systems \& Computers, 2003. vol.~2, pp. 1398--1402 (2003)

\bibitem{yang2014single}
Yang, C.Y., Ma, C., Yang, M.H.: Single-image super-resolution: A benchmark. In:
  European Conference on Computer Vision. pp. 372--386. Springer (2014)

\bibitem{yy2013}
Yang, C.Y., Yang, M.H.: Fast direct super-resolution by simple functions. In:
  Proceedings of the IEEE international conference on computer vision. pp.
  561--568 (2013)

\bibitem{ywhm2010}
Yang, J., Wright, J., Huang, T.S., Ma, Y.: Image super-resolution via sparse
  representation. IEEE transactions on image processing  \textbf{19}(11),
  2861--2873 (2010)

\bibitem{zhang2018learning}
Zhang, K., Zuo, W., Zhang, L.: Learning a single convolutional super-resolution
  network for multiple degradations. In: Proceedings of the IEEE Conference on
  Computer Vision and Pattern Recognition. pp. 3262--3271 (2018)

\bibitem{sr-sim}
Zhang, L., Li, H.: Sr-sim: A fast and high performance iqa index based on
  spectral residual. In: 2012 19th IEEE international conference on image
  processing. pp. 1473--1476. IEEE (2012)

\bibitem{vsi}
Zhang, L., Shen, Y., Li, H.: Vsi: A visual saliency-induced index for
  perceptual image quality assessment. IEEE Transactions on Image Processing
  \textbf{23}(10),  4270--4281 (2014)

\bibitem{rfsim}
Zhang, L., Zhang, L., Mou, X.: Rfsim: A feature based image quality assessment
  metric using riesz transforms. In: 2010 IEEE International Conference on
  Image Processing. pp. 321--324. IEEE (2010)

\bibitem{fsim}
Zhang, L., Zhang, L., Mou, X., Zhang, D.: Fsim: A feature similarity index for
  image quality assessment. IEEE transactions on Image Processing
  \textbf{20}(8),  2378--2386 (2011)

\bibitem{zhang2019making}
Zhang, R.: Making convolutional networks shift-invariant again. In:
  International Conference on Machine Learning. pp. 7324--7334 (2019)

\bibitem{zhang2018unreasonable}
Zhang, R., Isola, P., Efros, A.A., Shechtman, E., Wang, O.: The unreasonable
  effectiveness of deep features as a perceptual metric. In: Proceedings of the
  IEEE Conference on Computer Vision and Pattern Recognition. pp. 586--595
  (2018)

\bibitem{zhang2019ranksrgan}
Zhang, W., Liu, Y., Dong, C., Qiao, Y.: Ranksrgan: Generative adversarial
  networks with ranker for image super-resolution. In: Proceedings of the IEEE
  International Conference on Computer Vision. pp. 3096--3105 (2019)

\bibitem{rcan2018}
Zhang, Y., Li, K., Li, K., Wang, L., Zhong, B., Fu, Y.: Image super-resolution
  using very deep residual channel attention networks. In: Proceedings of the
  European Conference on Computer Vision (ECCV). pp. 286--301 (2018)

\end{thebibliography}
\end{document}